\documentclass[two column]{revtex4}%
\usepackage{amsfonts}
\usepackage{dsfont}
\usepackage{amsmath}
\usepackage{amssymb}
\usepackage{graphicx}%
\usepackage[unicode=true]{hyperref}
\providecommand{\U}[1]{\protect\rule{.1in}{.1in}}
\begin{document}
\title{St\"{u}ckelberg interferometry using spin-orbit-coupled cold atoms in an
optical lattice}
\author{Shuang Liang$^{1}$, Zheng-Chun Li$^{1}$, Weiping Zhang$^{2,4}$, Lu
Zhou$^{1,4}$\footnote{lzhou@phy.ecnu.edu.cn }, and Zhihao Lan$^{3}%
$\footnote{z.lan@ucl.ac.uk }}
\affiliation{$^{1}$State Key laboratory of Precision Spectroscopy, Department of Physics,
School of Physics and Electronic Science, East China Normal University,
Shanghai 200241, China}
\affiliation{$^{2}$Department of Physics and Astronomy, Shanghai Jiaotong University and
Tsung-Dao Lee Institute, Shanghai 200240, China}
\affiliation{$^{3}$Department of Electronic and Electrical Engineering, University College
London, Torrington Place, London, WC1E 7JE, United Kingdom}
\affiliation{$^{4}$Collaborative Innovation Center of Extreme Optics, Shanxi University,
Taiyuan, Shanxi 030006, China}

\begin{abstract}
Time evolution of spin-orbit-coupled cold atoms in an optical lattice is
studied, with a two-band energy spectrum having two avoided crossings. A force
is applied such that the atoms experience two consecutive Landau-Zener
tunnelings while transversing the avoided crossings. St\"{u}ckelberg
interference arises from the phase accumulated during the adiabatic evolution
between the two tunnelings. This phase is gauge field-dependent and thus
provides new opportunities to measure the synthetic gauge field, which is
verified via calculation of spin transition probabilities after a double passage
process. Time-dependent and time-averaged spin probabilities are derived, in
which resonances are found. We also demonstrate chiral Bloch oscillation and
rich spin-momentum locking behavior in this system.

\end{abstract}
\maketitle


\section{introduction}

\label{sec_introduction}

Atom interferometry has been proven to be a powerful tool for precision
metrology \cite{atom interferometry}. Besides that atom interferometry can be
used to test fundamental physical theory such as general relativity
\cite{kasevich2007,kasevich2008,muller2008,zoestScience,folmanScience}. Recent
years have also witnessed growing interest in atom interferometry with
synthetic gauge fields \cite{anderson2011,zhuPRA2012,jacob2007,lewenstein2005}%
, mostly due to the fact that synthetic gauge fields such as spin-orbit (SO)
coupling can be generated in neutral cold atoms via atom-light interaction
\cite{spielmanNature2011,celiPRL2014,gaugefieldReview}. The gauge field can
couple the atom's internal spin states to its center-of-mass motion, thus the
path difference in coordinate space can be mapped to a spin interference
signal in an interferometer setup, which can be used to measure external ac
force \cite{anderson2011} and demonstrate non-Abelian Aharonov-Bohm effect
\cite{jacob2007,lewenstein2005}. It is interesting to note that non-Abelian
gauge fields in real space have been observed recently \cite{nonabelian real
space} and a five-dimensional non-Abelian gauge field has also been explored
\cite{spielmanScience2018,spielman2019}.

In this work we concentrate on St\"{u}ckelberg interferometry. The principle
of St\"{u}ckelberg interference lies in a quantum system traveling through
more than one avoided crossing of two of its energy levels
\cite{landau,stuckelberg,shevchenkoPR2010,nakamuraBook}. Avoided crossing
represents a close encounter of two energy levels without actual degeneracy
\cite{haake}. Passing through the avoided crossings leads to Landau-Zener (LZ)
tunneling, which coherently splits and recombines wavefunction between the two
energy levels. Thus the phase accumulated between two LZ transitions
(sometimes referred as the St\"{u}ckelberg phase) leads to quantum
interference. The avoided crossings function as beam splitters in parameter
(quasi-momentum or time) space, in the sense that the physical principle
underlying St\"{u}ckelberg interferometry is identical to that of
Mach--Zehnder interferometer. St\"{u}ckelberg interferometry has been
implemented in molecule formation in ultracold atoms \cite{grimm2007}, atomic
Bose-Einstein condensates (BECs) in an optical lattice
\cite{klingPRL2010,zenesini2010}, electronic spins in a nitrogen-vacancy center
\cite{du2011} and recently acoustic modes \cite{mika2019}. The interband LZ
tunneling of ultracold atoms in bichromatic optical lattices
\cite{korsch2011,cai2011,yamakoshi2019} and between high energy bands
\cite{zhou2016,wimberger2012,plotzEPJD2011} are also predicted to have the
potential to perform St\"{u}ckelberg interference.

Recently a periodically driven SO-coupled atomic BECs in free space was used
to implement St\"{u}ckelberg interference \cite{chenPRA2017}. Using periodic
modulation of Raman coupling to create a pair of avoided crossings in the
energy dispersion, the resulting interference fringes become modulation
frequency dependent. Besides scenarios in free space, synthetic SO coupling
have also been implemented in ultracold atomic gas trapped in optical lattice
via either optical clock transition
\cite{fallaniPRL2016,wallPRL2016,kolkowitzNature}, double-well optical lattice
\cite{ketterlePRL2016,blochNP2014}, Raman-dressing
\cite{EngelsPRL2015,chenScience} and using a two-dimensional manifold of
momentum states \cite{gadwaySA}. By considering spin as a synthetic dimension,
these experiments can resemble the model of a two-leg ladder subject to a
magnetic flux \cite{hugelPRA2014}. Two-leg ladder has been realized using a
single optical cavity with two independent synthetic dimensions \cite{fan2019}
and the tunneling problem of a ladder system has also been addressed with
electronic setup \cite{takahashiJPSP2018,takahashi2018}. Many interesting
phenomena such as chiral Bloch oscillation, bandgap closing, edge state and
unconventional phases have been predicted in such a system by incorporating
diagonal couplings, additional legs, strong and long-range interactions
\cite{natuPRA2015,greschnerPRA2016,ghoshPRA2017,yangPRA2017,petrescuPRB2017,barbarinoPRA2018,citroPRB2018,haugQST2018,shinPRL2019,fallaniScience,haller2018,juneman2017,strinati2017,spielmanScience,greinerNature,tschischikPRA2015,dina2019}%
.

Motivated by these developments, in this work we consider a St\"{u}ckelberg
interferometry using SO-coupled cold atoms in an optical lattice. We found the
conditions to realize two avoided crossings in the energy dispersion of this
two-band system, which can be achieved in experiment and thus the system can
be used to perform St\"{u}ckelberg interference. The St\"{u}ckelberg
interferometry represents an atom interferometry with synthetic gauge fields
and provides new opportunity to study novel SO-coupled band structures. We
demonstrate that the interference pattern reveals a phase which depends on the
synthetic magnetic flux, from which the information on the synthetic gauge
field can be derived. Although the synthetic magnetic flux can be directly
probed via measuring the atomic SO momentum transfer, here St\"{u}ckelberg
interference provides an alternative way to demonstrate the effect of
synthetic gauge field without inquiring the atomic momentum information. The
time evolution is also explicitly studied, where resonances and chiral Bloch
oscillation are predicted.

The article is organized as follows: In Sec. \ref{sec_model} we present our
model and the effective Hamiltonian is derived. The principle of
St\"{u}ckelberg interference is illustrated in Sec. \ref{sec_stukelberg}. We
use the adiabatic-impulse model to analyze a double passage process in which
two avoided crossings are transversed. A more general discussion on the
dynamics is also performed with Floquet-Bloch theory. Sec. \ref{sec_chiral} is
devoted to the discussion of chiral Bloch oscillation in the case that no LZ
transitions take place and finally we conclude in Sec. \ref{sec_conclusion}.

\section{model}

\label{sec_model}

\begin{figure}[h]
\includegraphics[width=8cm]{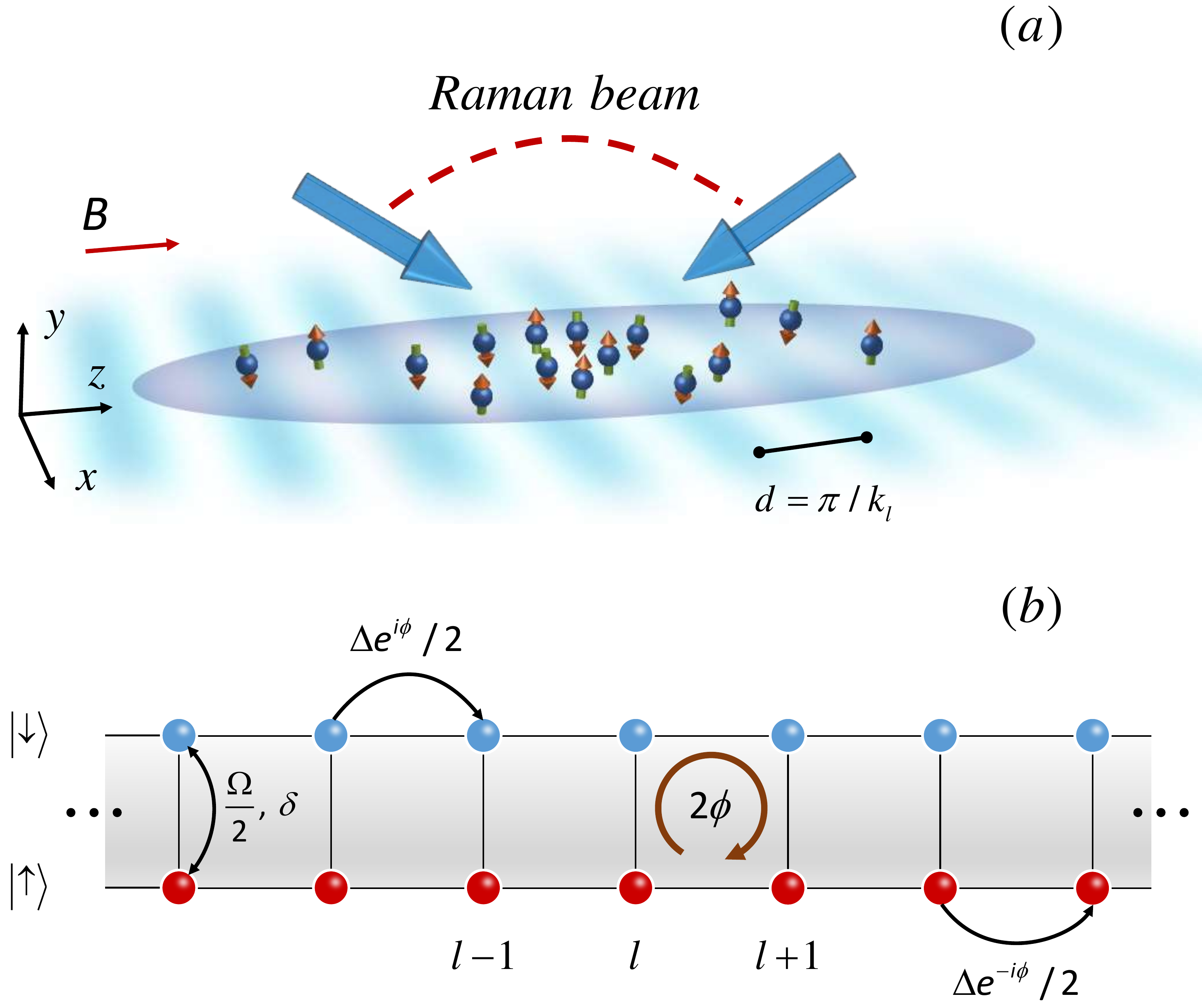}\caption{{\protect\footnotesize Schematic
diagram showing the system under consideration. (a) Setup: }$^{87}%
${\protect\footnotesize Rb cold atoms are confined in a one-dimensional
optical lattice along the }$z${\protect\footnotesize -direction, inside which
the effective SO interaction is induced via coupling the }$\left\vert
\downarrow\right\rangle $ {\protect\footnotesize and }$\left\vert
\uparrow\right\rangle $ {\protect\footnotesize hyperfine states with Raman
lasers. A bias magnetic field }$B$ {\protect\footnotesize causing quadratic
Zeeman splitting is applied along the }$z${\protect\footnotesize -direction.
(b) The hyperfine states can be treated as an effective synthetic dimension
made by two sites connected with a coherent tunneling, resulting in a two-leg
ladder pierced by a synthetic magnetic flux }$2\phi$
{\protect\footnotesize per plaquette.}}%
\label{fig_ladder}%
\end{figure}

In this work we consider the following model typically representing a two-leg
bosonic ladder pierced by a magnetic flux, which can be described by the
Hamiltonian ($\hbar=1$)%
\begin{align}
\hat{H}  &  =-\frac{\Delta}{2}\sum_{l}\left(  \hat{c}_{l}^{\dagger}%
e^{i\phi\hat{\sigma}_{z}}\hat{c}_{l+1}+\text{H.c.}\right) \nonumber\\
&  +\sum_{l}\hat{c}_{l}^{\dagger}\left(  \frac{\Omega}{2}\hat{\sigma}%
_{x}-\frac{\delta}{2}\hat{\sigma}_{z}\right)  \hat{c}_{l}.
\label{eq_hamiltonian_wannier}%
\end{align}
Hamiltonian (\ref{eq_hamiltonian_wannier}) can be implemented in the system of
Raman-dressed $^{87}$Rb cold atoms trapped in a one-dimensional optical
lattice \cite{EngelsPRL2015,celiPRL2014,zhouPRA2019} along the $z$-direction
under lowest energy band truncation and tight-binding approximation, as shown
in Fig. \ref{fig_ladder}(a). Here the atomic hyperfine states $\left\vert
1,-1\right\rangle $ ($\left\vert \downarrow\right\rangle $) and $\left\vert
1,0\right\rangle $ ($\left\vert \uparrow\right\rangle $) are coupled via Raman
lasers, thus generating effective SO interaction. By considering the atomic
pseudospin as a synthetic dimension, the system can be exactly mapped to a
two-leg bosonic ladder, as shown in Fig. \ref{fig_ladder}(b). The
spin-momentum locking in Raman transition is equivalent to spin-dependent
(leg-dependent) tunneling between neighboring sites, thus a particle hopping
around an elementary plaquette picks up an Aharonov-Bohm phase $2\phi$, which
is equivalent to the presence of an effective magnetic flux $2\phi$ per
plaquette piercing the system. $\phi=\pi k_{R}/k_{l}$ with $k_{R\left(
l\right)  }$ the wavevector of Raman beams projected onto the $z$-axis and the
laser forming the lattice, respectively. Here two-component annihilation
operator $\hat{c}_{l}=\left(  \hat{c}_{l\uparrow},\hat{c}_{l\downarrow
}\right)  ^{T}$ and $\Delta$ is half-bandwidth on the scale of kHz in typical
experiments with $^{87}$Rb atoms \cite{estimate}. The inter-leg coupling is
characterized by the hopping amplitude $\Omega/2$ and detuning $\delta
$.

Compared with the previously studied two-leg ladder \cite{hugelPRA2014}, here
the inter-leg detuning $\delta$ is additionally taken into account. We note
that the two-photon detuning $\delta$ is usually available via bias magnetic
field in typical experiments with Raman-dressed BECs
\cite{EngelsPRL2015,spielmanNature2011,ketterlePRL2016}. Making the Fourier
transform $\hat{c}_{q}=\sqrt{\frac{d}{2\pi}}\sum_{l}\hat{c}_{l}e^{-iqld}$ with
$\hat{c}_{q}=\left(  \hat{c}_{q\uparrow},\hat{c}_{q\downarrow}\right)  ^{T}$
and $d=\pi/k_{l}$ the lattice constant, which is equivalent to the transform
of the system from the Wannier basis to the Bloch basis, Hamiltonian
(\ref{eq_hamiltonian_wannier}) can be rewritten in the quasi-momentum basis as
$\hat{H}=\sum_{q}\hat{c}\hat{\mathcal{H}}_{q}\hat{c}_{q}$ with%
\begin{align}
\hat{\mathcal{H}}_{q}  &  =-\Delta\cos\phi\cos\left(  qd\right)
\hat{\mathds{1}}+\frac{\Omega}{2}\hat{\sigma}_{x}\nonumber\\
&  +\left[  -\frac{\delta}{2}+\Delta\sin\phi\sin\left(  qd\right)  \right]
\hat{\sigma}_{z}. \label{eq_hamiltonian q}%
\end{align}
Hamiltonian $\hat{\mathcal{H}}_{q}$ indicates a two-band structure with
$\varepsilon_{\pm}\left(  q\right)  =$ $-\Delta\cos\phi\cos\left(  qd\right)
\pm\sqrt{\left[  -\delta/2+\Delta\sin\phi\sin\left(  qd\right)  \right]
^{2}+\left(  \Omega/2\right)  ^{2}}$. When $\left\vert \delta/2\Delta\sin
\phi\right\vert <1$, there are two avoided crossings (at which spacing between
$\varepsilon_{\pm}$ takes minimal value $\Omega$) located in the first
Brillouin zone, thus making the system ideal for implementing
Landau-Zener-St\"{u}ckelberg interferometry.

\section{St\"{u}ckelberg interferometry}

\label{sec_stukelberg}

\begin{figure}[h]
\includegraphics[width=8cm]{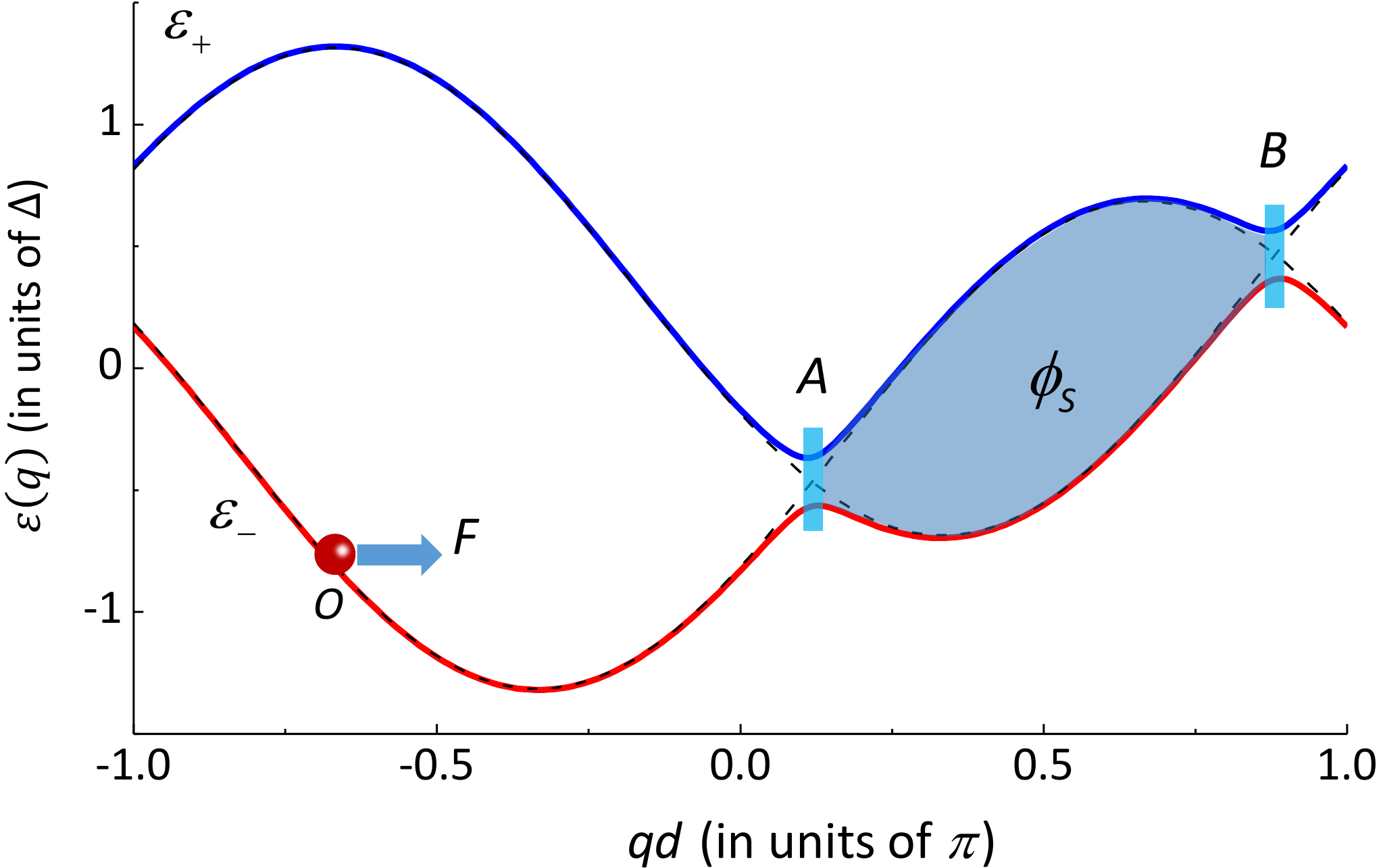}\caption{{\protect\footnotesize Energy band
}$\varepsilon\left(  q\right)  $ {\protect\footnotesize at }$\delta
=0.63\Delta${\protect\footnotesize , }$\Omega=0.2\Delta$
{\protect\footnotesize and }$\phi=\pi/3${\protect\footnotesize . The diabatic
energy levels are shown by dashed lines. This figure illustrates the principle
of two-leg ladder as a St\"{u}ckelberg interferometry.}}%
\label{fig_band}%
\end{figure}

The principle of St\"{u}ckelberg interferometry is illustrated in Fig.
\ref{fig_band}, where we assume $\delta/2\Delta\sin\phi>0$ without loss of
generality. The two avoided crossings then locate at $A$ with $q_{A}%
d=\arcsin\left(  \delta/2\Delta\sin\phi\right)  $ and $B$ with $q_{B}%
d=\pi-\arcsin\left(  \delta/2\Delta\sin\phi\right)  $. Suppose that the system
is initially prepared in the state $\left\vert q_{0},\uparrow\right\rangle $
with $q_{0}d\in\left[  -\pi,\arcsin\left(  \delta/2\Delta\sin\phi\right)
\right]  $ (for example $q_{0}$ could be the quasimomentum of the system
ground state), which is marked as point $O$ in Fig. \ref{fig_band}.
Furthermore, a constant external force $F$ is exerted on the atoms via tilting
the optical lattice, which drives Bloch oscillation of the atoms. Bloch
oscillation depicts the traveling of atoms along the energy bands. Upon
reaching the avoided crossing $A$,\ the atoms will coherently split into two
components via LZ transition. The two components then separately travel along
$\varepsilon_{\pm}\left(  q\right)  $ and thus acquire a different phase.
Finally the two components recombine and interfere with each other after
another LZ transition at the avoided crossing $B$, from which the information
on the phase difference $\phi_{S}$ accumulated during traversing
$\varepsilon_{\pm}\left(  q\right)  $ between $A$ and $B$ can be derived.

In the following we first study this double passage process (the avoided
crossing region is passed twice) using adiabatic-impulse model in Sec.
\ref{sec_double passage}, where a simple relation between spin population and
St\"{u}ckelberg phase is found. This model can help us gain physical insight
into the St\"{u}ckelberg interferometry. Then the spin dynamics is studied in
Sec. \ref{sec_average occupation}, from which one can calculate the
time-averaged spin population. This is relevant to experiment and the
corresponding interference patterns will be identified.

\subsection{Double passage: Adiabatic-impulse model}

\label{sec_double passage}

Under the assumption that $F$ is weak enough not to induce interband
transitions, the adiabatic approximation can be applied, under which the atoms
move adiabatically along the energy band with the quasimomentum $q\left(
t\right)  =q_{0}+Ft$ \cite{castinPRA1997}\ except in the vicinity of the
avoided crossings. The non-adiabatic evolution of atoms while traversing the
avoided crossing region is considered to take place instantaneously, which
validates at $\Omega^{2}+\left(  2\Delta\sin\phi\right)  ^{2}>>\left(
Fd\right)  ^{2}$ \cite{garraway1997}. The double passage process from
$t_{i}=t_{A}^{-}$ to $t_{f}=t_{B}^{+}$ with $t_{A\left(  B\right)  }=\left(
q_{A\left(  B\right)  }-q_{0}\right)  /F$ can be described by a transfer
matrix $\mathcal{T}_{D}$ in the diabatic basis (bare spin basis) as%
\begin{equation}
c\left(  t_{f}\right)  =\mathcal{T}_{D}c\left(  t_{i}\right)  ,
\label{eq_transfer}%
\end{equation}
where $c\left(  t\right)  =\left\langle \hat{c}_{q}\right\rangle =\left(
c_{\uparrow},c_{\downarrow}\right)  ^{T}$ represents atomic population
amplitude in the diabatic basis (bare spin basis) and governed by the
equations of motion $idc\left(  t\right)  /dt=\hat{\mathcal{H}}_{q}c\left(
t\right)  $. In the adiabatic-impulse approximation \cite{damskiPRA2006},
$\mathcal{T}_{D}$ can be\ devided into 3 parts:

\romannumeral1. The LZ transition at the avoided crossing $A$. To symmetrize
the two diabatic energy levels, we treat $c\left(  t\right)  \exp\left\{
i\Delta\cos\phi\int_{0}^{t}dt^{\prime}\cos\left[  q\left(  t^{\prime}\right)
d\right]  \right\}  $ as the wavevector, submit it into the equations of
motion and can have%

\begin{equation}
i\frac{dc_{\uparrow\left(  \downarrow\right)  }}{dt}=\frac{\Omega}%
{2}c_{\downarrow\left(  \uparrow\right)  }\pm\left[  -\frac{\delta}{2}%
+\Delta\sin\phi\sin\left(  qd\right)  \right]  c_{\uparrow\left(
\downarrow\right)  }. \label{eq_LZ1}%
\end{equation}
In the vicinity of $A$ with $\left\vert Fd\left(  t-t_{A}\right)  \right\vert
<<1$, Eq. (\ref{eq_LZ1}) can be linearized as%
\begin{equation}
i\frac{dc_{\uparrow\left(  \downarrow\right)  }}{dt^{\prime}}=\frac{\Omega}%
{2}c_{\downarrow\left(  \uparrow\right)  }\pm\frac{v}{2}t^{\prime}%
c_{\uparrow\left(  \downarrow\right)  } \label{eq_LZ2}%
\end{equation}
with $v=2\Delta Fd\sin\phi\cos\left(  q_{A}d\right)  =2\Delta Fd\sin\phi
\sqrt{1-\left(  \delta/2\Delta\sin\phi\right)  ^{2}}$ and $t^{\prime}=t-t_{A}%
$. Eq. (\ref{eq_LZ2}) defines the standard LZ problem for which the exact
solution can be expressed in terms of parabolic cylinder functions
\cite{landau,shevchenkoPR2010}. Then the nonadiabatic transition is described
by $\left[  c_{\uparrow}\left(  +0\right)  ,c_{\downarrow}\left(  +0\right)
\right]  ^{T}=\mathcal{T}_{A}\left[  c_{\uparrow}\left(  -0\right)
,c_{\downarrow}\left(  -0\right)  \right]  ^{T}$ with the time-independent
matrix%
\begin{equation}
\mathcal{T}_{A}=\left(
\begin{array}
[c]{cc}%
\sqrt{P_{LZ}} & \sqrt{1-P_{LZ}}e^{-i\varphi_{st}}\\
-\sqrt{1-P_{LZ}}e^{i\varphi_{st}} & \sqrt{P_{LZ}}%
\end{array}
\right)  , \label{eq_ta}%
\end{equation}
where the LZ transition probability $P_{LZ}=\exp\left(  -2\pi\xi\right)  $ and
the Stokes phase $\varphi_{st}=\pi/4+\xi\left(  \ln\xi-1\right)  +\arg
\Gamma\left(  1-i\xi\right)  $, with $\xi=\Omega^{2}/4v$ and the gamma
function $\Gamma$.

\romannumeral2. The adiabatic evolution in the region between the two avoided
crossings $A$ and $B$. One can notice that in this region far from the avoided
crossings the diabatic energy for spin-$\uparrow$($\downarrow$) components
coincides with the energy dispersion $\varepsilon_{\pm}\left(  q\right)  $
respectively, i.e., the excited eigenstate of $\hat{\mathcal{H}}_{q}$ coincide
with spin-$\uparrow$ while the ground eigenstate coincide with
spin-$\downarrow$, as shown in Fig. \ref{fig_band}. This enables us to write
the evolution matrix as%
\begin{equation}
\mathcal{U}=\left(
\begin{array}
[c]{cc}%
e^{-i\frac{\phi_{S}}{2}} & 0\\
0 & e^{i\frac{\phi_{S}}{2}}%
\end{array}
\right)  \label{eq_u}%
\end{equation}
with $\phi_{S}=\int_{q_{A}}^{q_{B}}dq\left[  \varepsilon_{+}\left(  q\right)
-\varepsilon_{-}\left(  q\right)  \right]  /F$.

\romannumeral3. The LZ transition at the avoided crossing $B$. This process is
identical to \romannumeral1 \ except that $v\rightarrow-v$ in Eq.
(\ref{eq_LZ2}). Then the LZ transition matrix $\mathcal{T}_{B}=\mathcal{T}%
_{A}^{T}$.

Combine the process \romannumeral1-\romannumeral3, the transfer matrix
$\mathcal{T}_{D}$ has the form%
\begin{align}
\mathcal{T}_{D}  &  =\mathcal{T}_{B}\mathcal{UT}_{A}=\left(
\begin{array}
[c]{cc}%
\alpha & \beta\\
\beta & \alpha^{\ast}%
\end{array}
\right)  ,\nonumber\\
\alpha &  =P_{LZ}e^{-i\frac{\phi_{S}}{2}}+\left(  1-P_{LZ}\right)  e^{i\left(
\frac{\phi_{S}}{2}+\varphi_{st}\right)  },\nonumber\\
\beta &  =-2i\sqrt{P_{LZ}\left(  1-P_{LZ}\right)  }\sin\left(  \frac{\phi_{S}%
}{2}+\varphi_{st}\right)  . \label{eq_td}%
\end{align}
Eqs. (\ref{eq_transfer}) and (\ref{eq_td}) indicate that the transition
probability from spin-$\uparrow$ to $\downarrow$ is $\left\vert \beta
\right\vert ^{2}$, which oscillates with the phase $\phi_{S}/2+\varphi_{st}$.
Since $\phi_{S}\propto1/F$, then under adiabatic approxiation with small force
$F$ the St\"{u}ckelberg phase $\phi_{S}$ is dominant and the contribution from
$\varphi_{st}$ can be neglected \cite{takahashiJPSP2018}, i.e., one can
neglect the quantum phase and focus on the semiclassical lattice
effect.\begin{figure}[h]
\includegraphics[width=8cm]{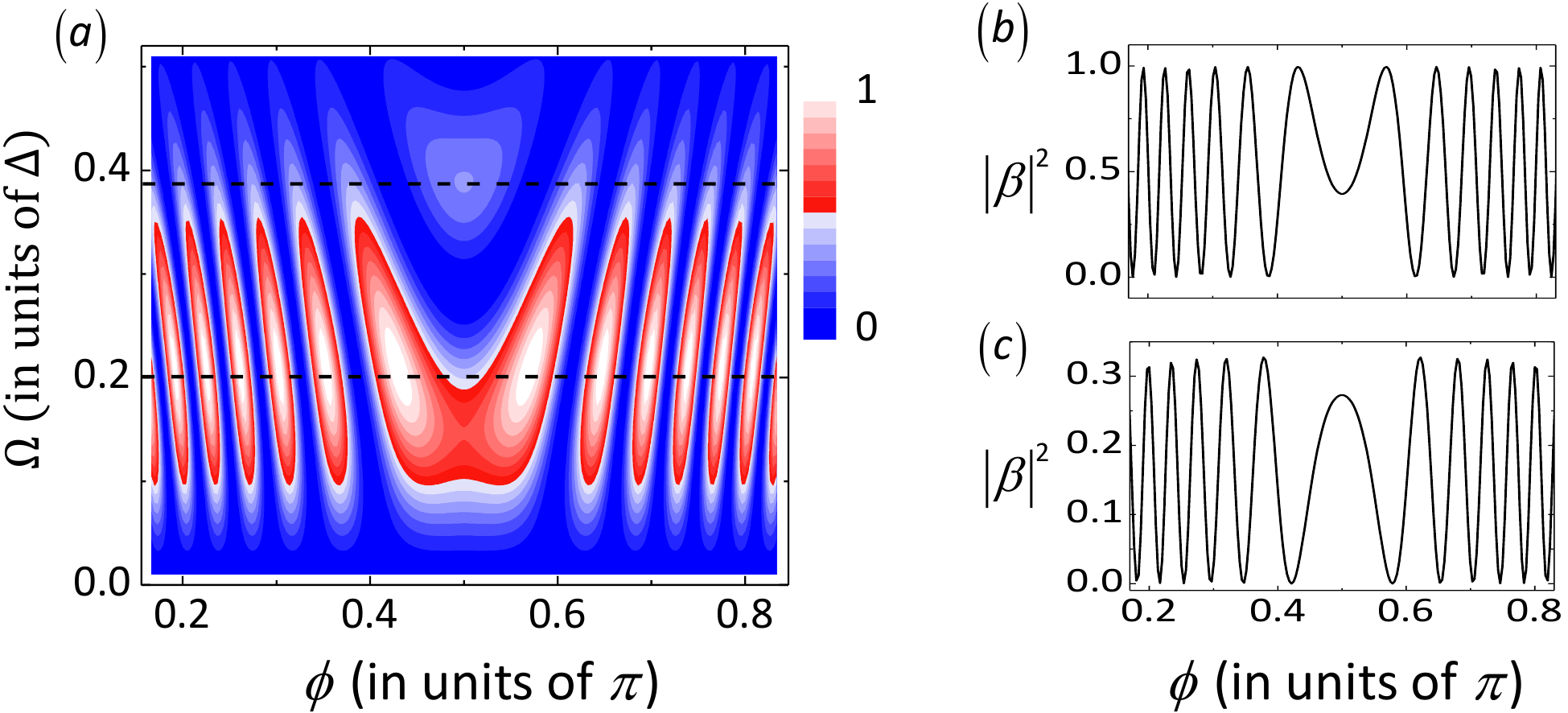}\caption{{\protect\footnotesize (a) Contour
plot of the transition probability }$\left\vert \beta\right\vert ^{2}$
{\protect\footnotesize in the }$\left(  \phi\text{, }\Omega\right)  $
{\protect\footnotesize parameter plane with }$\delta=0.23\Delta$
{\protect\footnotesize and }$Fd=0.05\Delta${\protect\footnotesize . The dashed
lines mark }$\Omega=0.2\Delta$ {\protect\footnotesize and }$0.39\Delta$
{\protect\footnotesize respectively with the corresponding oscillations shown
in (b) and (c).}}%
\label{fig_tran prob}%
\end{figure}

We calculate the transition probability $\left\vert \beta\right\vert ^{2}$ as
a function of the synthetic magnetic flux $\phi$ and the Raman coupling
strength $\Omega$ with the results shown in Fig. \ref{fig_tran prob}. This
figure is symmetric with respect to $\phi=\pi/2$ as $\left\vert \beta
\right\vert ^{2}$ is a function of $\sin\phi$. On both sides of $\phi=\pi/2$
it displays approximately periodic oscillation which can be recognized as
St\"{u}ckelberg oscillation. The appearance of St\"{u}ckelberg oscillation
versus $\phi$ reflects the nearly monotonic dependence of the St\"{u}ckelberg
phase $\phi_{S}$ on $\phi$ since the value of $P_{LZ}$ is insensitive to the
variation of $\phi$. The interval between two neighboring maximal(minimal)
transition probabilities is then given by $\delta\phi_{S}\approx2\pi$. The
calculation is performed with $Fd=0.05\Delta$\ and the oscillation will become
stronger for a smaller force $F$. For the parameters considered here, at
$\Omega=0.2\Delta$ we have $P_{LZ}\approx0.5$ and hence the transition
probability can reach the maxima $1$, as shown in Fig. \ref{fig_tran prob}(b).
Since $\Omega$ determines the energy spacing at the avoided crossings and the
LZ transition probability $P_{LZ}$ is exponentially dependent on it, one can
then generally expect that with increasing deviation from $\Omega=0.2\Delta$
the transition probability $\left\vert \beta\right\vert ^{2}$ will gradually
decrease. This is the case except for the region around $\phi=\pi/2$. As
$\phi_{S}$ is a complex function of $\Omega$, another local maximum of the
transition probability appears around $\Omega=0.39\Delta$, as shown in Fig.
\ref{fig_tran prob}(c).

By measuring St\"{u}ckelberg interference fringes one can map out the novel
band structure with SO coupling \cite{klingPRL2010,chenPRA2017}. As compared
with the experiment \cite{chenPRA2017}, here the Raman coupling is not
periodically modulated. However in the scenario with optical lattice, it can
take the effect as to engineer the dispersion and realize two avoided
crossings, thus making the St\"{u}ckelberg interference feasible. The
interference fringes in \cite{chenPRA2017} rely on the frequency and amplitude
of periodical modulation applied on the Raman beams, showing the effect of
Floquet engineering on SO-coupled bands. Here the physics underlying
St\"{u}ckelberg interference is the presence of synthetic magnetic flux. In
the meanwhile considering the fact that the centre of interference (or zeroth
order fringe corresponding to $\phi=\pi/2$) can be easily identified due to
the symmetric nature of the interference pattern shown in Fig.
\ref{fig_tran prob}, it enables one to recognize the order of interference
fringe which can be used to indicate the synthetic magnetic flux $\phi$. In
this sense the St\"{u}ckelberg interferometer demonstrated here also provides
a new opportunity to measure the synthetic gauge field.

\subsection{Dynamics: Floquet-Bloch theory}

\label{sec_average occupation}

The dynamics considered here is mathmatically similar to Bloch oscillation in
a two-band system with interband coupling, however many previous works have
dealt with this kind of problem \cite{rotvigPRL1995,zhaoPRB1996} without SO
coupling. Here we perform the analysis in terms of the Floquet-Bloch operator
\cite{kolovskyPRE2003,plotzEPJD2011}. We can start from Hamiltonian
(\ref{eq_hamiltonian_wannier}) with an additional term $-Fd\sum_{l}l\hat
{c}_{l}^{\dagger}\hat{c}_{l}$ describing that the lattice is tilted with
on-site energies $-lFd$. This term can be removed via making a unitary
transform with the operator $\exp\left(  iFdt\sum_{l}l\hat{c}_{l}^{\dagger
}\hat{c}_{l}\right)  $. Then following the same process as that has been
performed in Sec. \ref{sec_model}, we obtain the resulting periodic
Hamiltonian in the Bloch basis as%
\begin{align}
\hat{\mathcal{H}}_{q}^{\prime}\left(  t\right)   &  =-\Delta\cos\phi
\cos\left[  \left(  q+Ft\right)  d\right]  \hat{\mathds{1}}+\frac{\Omega}%
{2}\hat{\sigma}_{x}\nonumber\\
&  +\left\{  -\frac{\delta}{2}+\Delta\sin\phi\sin\left[  \left(  q+Ft\right)
d\right]  \right\}  \hat{\sigma}_{z}.\label{eq_hq1}%
\end{align}
Introducing $c_{\uparrow\left(  \downarrow\right)  }\left(  q,t\right)
=\tilde{c}_{\uparrow\left(  \downarrow\right)  }\left(  q,t\right)
\exp\left\{  \pm i\delta t/2+i\Delta\int_{0}^{t}dt^{\prime}\cos\left[  \left(
q+Ft^{\prime}\right)  d\pm\phi\right]  \right\}  $ to remove the diagonal
terms in the equations of motion, we have%
\begin{equation}
i\frac{\partial}{\partial t}\binom{\tilde{c}_{\uparrow}\left(  q,t\right)
}{\tilde{c}_{\downarrow}\left(  q,t\right)  }=\frac{\Omega}{2}\left[
\cos\left(  \phi_{D}\right)  \hat{\sigma}_{x}+\sin\left(  \phi_{D}\right)
\hat{\sigma}_{y}\right]  \binom{\tilde{c}_{\uparrow}\left(  q,t\right)
}{\tilde{c}_{\downarrow}\left(  q,t\right)  },\label{eq_motion}%
\end{equation}
where $\phi_{D}\left(  q,t\right)  =\delta t+\left(  2\Delta\sin
\phi/Fd\right)  \left\{  \cos\left[  \left(  q+Ft\right)  d\right]
-\cos\left(  qd\right)  \right\}  $ is the dynamical phase between the two
legs. In the case of weak interleg coupling $\Omega$ (which is indeed the case
we are studying), one can have%
\begin{align}
\binom{\tilde{c}_{\uparrow}\left(  q,t\right)  }{\tilde{c}_{\downarrow}\left(
q,t\right)  } &  =\exp\left\{  -i\frac{\Omega}{2}\int_{0}^{t}dt^{\prime
}\left[  \cos\left[  \phi_{D}\left(  q,t^{\prime}\right)  \right]  \hat
{\sigma}_{x}\right.  \right.  \nonumber\\
&  \left.  \left.  +\sin\left[  \phi_{D}\left(  q,t^{\prime}\right)  \right]
\hat{\sigma}_{y}\right]  \right\}  \binom{\tilde{c}_{\uparrow}\left(
q,0\right)  }{\tilde{c}_{\downarrow}\left(  q,0\right)  },\label{eq_magnus}%
\end{align}
which can be recognized as Magnus expansion to the first order
\cite{plotzEPJD2011}. For the atoms initially prepared in spin-$\uparrow$ leg,
the spin transition probability is given by $P_{\downarrow}=\sin^{2}\left(
\Omega\left\vert \chi\right\vert /2\right)  $ with%
\begin{align}
\chi &  =\int_{0}^{t}dt^{\prime}e^{i\phi_{D}\left(  q,t^{\prime}\right)
}\nonumber\\
&  =2\sum_{n}J_{n}\left(  z\right)  e^{i\delta_{n}t+in\left(  qd+\pi/2\right)
-iz\cos\left(  qd\right)  }\frac{\sin\left(  \delta_{n}t/2\right)  }%
{\delta_{n}},\label{eq_chi}%
\end{align}
where $z=2\left(  \Delta/Fd\right)  \sin\phi$, $\delta_{n}=\delta+nFd$ and
$J_{n}\left(  z\right)  $ is the $n$-th order Bessel function of the first
kind. Without loss of generality here we assume $qd=-\pi/2$, which can always
be achieved by shifting the starting time. Then the transition probability
$P_{\downarrow}$ reads%
\begin{equation}
P_{\downarrow}\left(  t\right)  =\sin^{2}\left[  \Omega\left\vert \sum
_{n}J_{n}\left(  z\right)  e^{i\delta_{n}t}\frac{\sin\left(  \delta
_{n}t/2\right)  }{\delta_{n}}\right\vert \right]  .\label{eq_pd}%
\end{equation}
At some specific values of $n=m$ with $\delta_{m}\sim0$ an interleg coupling
is on resonance, then Eq. (\ref{eq_pd}) becomes%
\begin{align}
P_{\downarrow}\left(  t\right)   &  =\sin^{2}\left[  \Omega\left\vert
J_{m}\left(  z\right)  \frac{t}{2}\right.  \right.  \nonumber\\
&  \left.  \left.  +\sum_{n\neq m}J_{n}\left(  z\right)  e^{i\delta_{n}t}%
\frac{\sin\left(  \delta_{n}t/2\right)  }{\delta_{n}}\right\vert \right]
,\label{eq_pd1}%
\end{align}
which typically represents a large period (determined by $2\pi/\Omega
J_{m}\left(  z\right)  $) oscillation with small-amplitude high-frequency
oscillations superimposed on it. Eq. (\ref{eq_pd1}) indicates that in long run
the first term will dominate over other interleg couplings, thus the averaged
probability $\overline{P_{\downarrow}}=1/2$. In the case that no resonance
takes place and in the meanwhile if $z\leq1$, the $n=0$ term will dominate and
then one can have%
\begin{align}
P_{\downarrow}\left(  t\right)   &  \approx\sin^{2}\left[  \frac{\Omega
J_{0}\left(  z\right)  }{\delta}\sin\left(  \delta t/2\right)  \right]
\nonumber\\
&  =\left\{  2\sum_{n=1}^{\infty}J_{2n-1}\left(  V_{0}\right)  \sin\left[
\left(  n-\frac{1}{2}\right)  \delta t\right]  \right\}  ^{2},\label{eq_pd2}%
\end{align}
which can be approximately reduced to ${V}_{0}^{2}\sin^{2}\left(  \delta
t/2\right)  $ if $\left\vert V_{0}\right\vert =\left\vert \Omega J_{0}\left(
z\right)  /\delta\right\vert <<1$. This will lead to an averaged probability
of $\overline{P_{\downarrow}}=V_{0}^{2}/2$. Combining the discussions above,
one can understand that the averaged transition probability $\overline
{P_{\downarrow}}$ will vanish at large values of $\delta$ unless resonances
take place at $\delta=nFd$.\begin{figure}[h]
\includegraphics[width=8cm]{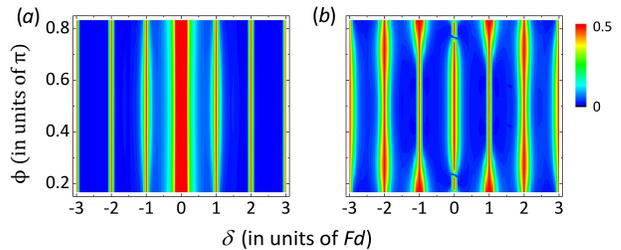}\caption{{\protect\footnotesize Contour plot of
the averaged probability }$\overline{P_{\downarrow}}$%
{\protect\footnotesize \ in the }$\left(  \delta\text{, }\phi\right)
${\protect\footnotesize \ parameter plane with }$\Omega=0.2Fd$%
{\protect\footnotesize \ and (a) }$\Delta=0.5Fd${\protect\footnotesize ; (b)
}$\Delta=2Fd${\protect\footnotesize . The numerical simulations are performed
with Eq. (\ref{eq_pd}).}}%
\label{fig_averaged probability}%
\end{figure}

We plot the averaged probability $\overline{P_{\downarrow}}$ as a function of
$\delta$ and $\phi$ in Fig. \ref{fig_averaged probability} based on Eq.
(\ref{eq_pd}). Fig. \ref{fig_averaged probability}(a) is for $z<1$ while Fig.
\ref{fig_averaged probability}(b) for $z>1$. In both cases the resonances
around $\delta=nFd$ can be clearly observed, indicating the system can be used
for force detection. The precision of force detection is related to the
uncertainty of two-photon detuning $\delta$, which can be restricted, for
example, via performing Pound-Drever-Hall laser frequency stabilization on the
Raman beams. Fig. \ref{fig_averaged probability}(a) shows a weak dependence of
$\overline{P_{\downarrow}}$ on $\phi$, however at $z>1$ one can still expect
differences while varying $\phi$, as displayed in Fig.
\ref{fig_averaged probability}(b). In this figure one can even observe that on
the $\delta=0$ resonance $\overline{P_{\downarrow}}$ vanishes at some specific
values of $\phi$, which correspond to the points with $J_{0}\left(  z\right)
=0$.

\section{chiral bloch oscillation}

\label{sec_chiral}

In the case of weak force $F$ and large interleg coupling $\Omega$, i.e., no
interband transitions, the traditional Bloch oscillation takes place in the
present system with a period $T_{B}=2\pi/Fd$ and an amplitude propotional to
the bandwidth \cite{KartashovPRL2016,zhouPRA2019}. For the Hamiltonian
(\ref{eq_hamiltonian q}), at zero detuning $\delta=0$ the eigenstate of lowest
(highest) energy band has negative (positive) chirality. The chirality can be
defined as $\mathcal{C}=q\left(  \left\langle \hat{c}_{q\uparrow}^{\dagger
}\hat{c}_{q\uparrow}\right\rangle -\left\langle \hat{c}_{q\downarrow}%
^{\dagger}\hat{c}_{q\downarrow}\right\rangle \right)  =q\left\langle
\hat{\sigma}_{z}\right\rangle $. Then in the approximation under which the
atoms move adiabatically along the energy band, the Bloch oscillation will
exhibit chiral characteristics \cite{yangPRA2017}. For an atomic wavepacket
moving along the lowest band with negative chirality, the positive (negative)
momentum will tend to associate with spin-$\downarrow$($\uparrow$) atoms,
signaling spin-momentum locking. In addition to that, at $\delta=0$ and small
interleg coupling $\Omega$, the lowest energy band can have two energy
degenerate minima, characteristic of spin-orbit coupled systems. Increasing
$\Omega$ beyond a critical value, the two minima will merge into one minimum,
signaling a quantum phase transition from the vortex phase to the Meissner
phase, which was experimentally observed in \cite{blochNP2014}. This band
curvature change upon the phase transition can also be captured via Bloch
oscillation \cite{yangPRA2017}.\begin{figure}[h]
\includegraphics[width=8cm]{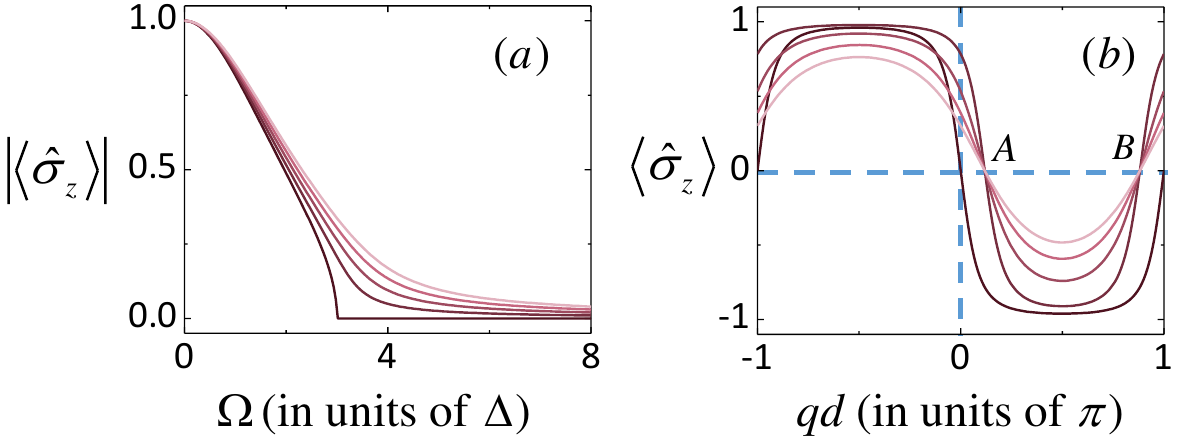}\caption{{\protect\footnotesize (a)
}$\left\vert \left\langle \hat{\sigma}_{z}\right\rangle \right\vert $
{\protect\footnotesize of the ground state as a function of }$\Omega
${\protect\footnotesize . From dark to light }$\delta=0$%
{\protect\footnotesize , }$0.05\Delta${\protect\footnotesize , }$0.1\Delta
${\protect\footnotesize , }$0.15\Delta${\protect\footnotesize , }$0.2\Delta
${\protect\footnotesize , respectively. (b) }$\left\langle \hat{\sigma}%
_{z}\right\rangle $ {\protect\footnotesize of the lowest band. The darkest
line is for }$\delta=0$ {\protect\footnotesize and }$\Omega=0.5\Delta
${\protect\footnotesize , indicating the eigenvectors in the whole band have
negative chirality. The other lines from dark to light are for }%
$\delta=0.63\Delta$ {\protect\footnotesize and }$\Omega=0.5\Delta
${\protect\footnotesize , }$\Delta${\protect\footnotesize , }$1.5\Delta
${\protect\footnotesize , }$2\Delta${\protect\footnotesize . The synthetic
magnetic flux }$\phi=\pi/3${\protect\footnotesize .}}%
\label{fig_chirality}%
\end{figure}

At any finite $\delta$, the energy band is asymmetric with respect to
inversion in Brillouin zone ($q\rightarrow-q$), which breaks time reversal
symmetry. As a result of time reversal symmetry breaking, Kramers degeneracy
does not hold here. However the lowest energy band can still exhibit two
nondegenerate local minima, typical of SO coupled systems as those have been
demonstrated in \cite{EngelsPRL2015}. In Fig. \ref{fig_chirality}(a), where
$\left\vert \left\langle \hat{\sigma}_{z}\right\rangle \right\vert $ of the
ground state is plotted as a function of $\Omega$, the first-order derivative
discontinuity disappears at any finite $\delta$, indicating that there exists
no quantum phase transition. In the meanwhile, chirality is not always
negative for all the eigenstates in the lowest energy band, as shown in Fig.
\ref{fig_chirality}(b). For the parameters considered here, at finite detuning
one can have negative chirality in the quasi-momentum region $\left[
-\pi\text{, }0\right]  $ and $\left[  q_{A}\text{, }q_{B}\right]  $ where
$q_{A\left(  B\right)  }$ are the same as those have been defined in Sec.
\ref{sec_stukelberg}, while in the rest region of the 1st Brillouin zone the
chirality becomes positive for ground eigenstates. This chirality change will
be reflected in the Bloch oscillation dynamics. \begin{figure}[h]
\includegraphics[width=8cm]{chirality1}\caption{{\protect\footnotesize Bloch
oscillation dynamics of }$\left\vert \psi_{l,\uparrow}\right\vert
^{2}+\left\vert \psi_{l,\downarrow}\right\vert ^{2}$
{\protect\footnotesize (top row), }$\left\vert \psi_{l,\uparrow}\right\vert
^{2}-\left\vert \psi_{l,\downarrow}\right\vert ^{2}$
{\protect\footnotesize (middle row) and }$\sum_{l}\left\vert \psi_{l,\sigma
}\right\vert ^{2}$ {\protect\footnotesize (bottom row). The left column
corresponds to }$\delta=0$ {\protect\footnotesize while the right column
corresponds to }$\delta=0.63\Delta${\protect\footnotesize . The other
parameters are set as }$\Omega=2\Delta${\protect\footnotesize , }$\phi=\pi/3$
{\protect\footnotesize and }$Fd=0.08\Delta${\protect\footnotesize . The
initial state is given by Eq. (\ref{eq_initial}) with }$w=5$%
{\protect\footnotesize , }$l_{0}=0$ {\protect\footnotesize and }$q_{0}%
=0${\protect\footnotesize .}}%
\label{fig_chiral bloch}%
\end{figure}

We simulate the Bloch oscillation dynamics using the method of eigenstate
expansion developed in \cite{zhouPRA2019}. Initially the atoms are assumed to
be prepared in a state $\left\vert \psi\right\rangle =\sum_{l,\sigma}%
\psi_{l,\sigma}\hat{c}_{l,\sigma}^{\dagger}\left\vert 0\right\rangle $ with%
\begin{equation}
\psi_{l}\left(  t=0\right)  =\frac{1}{\left(  2w\sqrt{\pi}\right)  ^{1/2}%
}e^{-\left(  l-l_{0}\right)  ^{2}/2w^{2}+iq_{0}ld}\binom{1}{1},
\label{eq_initial}%
\end{equation}
representing a spin-balanced Gaussian wave packet with width $w$, center
$l_{0}$ and initial quasi-momentum $q_{0}$. The results are shown in Fig.
\ref{fig_chiral bloch}. The left column is for $\delta=0$. The Bloch
oscillation in the top row clearly reflects the two degenerate minima nature
of the lowest band in the regime of the vortex phase. In the meanwhile the
middle and bottom rows also show clear evidence of chiral Bloch oscillation
and spin-momentum locking. The case of nonzero $\delta$ is shown in the right
column. Despite that the initial state (\ref{eq_initial}) is not exactly the
eigenstate, the atoms can still adiabatically follow the lowest band and
exhibit its curvature with two local minima. This process is associated with
chirality change, as shown in the middle and bottom rows. Spin-momentum
locking still exists, however it becomes nonmonotonic: The spin-$\uparrow
$($\downarrow$) atoms don't simply associate with negative (positive)
momentum, but now they are associated with certain momentum range, which can
be tuned via varying the detuning $\delta$.

\section{summary and outlook}

\label{sec_conclusion}

We note that physically the Raman lasers inducing SO interaction also
inevitably couple atoms to high-lying bands, which will affect the
single-particle physics \cite{cui2015,pan2016} such as dynamical instability
\cite{EngelsPRL2015}. This heating problem can be circumvented with an optical
clock transition \cite{fallaniPRL2016,wallPRL2016,kolkowitzNature}, which
connects the long lifetime states with a momentum transfer. One can perform a
gauge transformation in the quasi-momentum space and work with
quasi-momentum-shifted band structure. Taking atom collisions into account
will lead to\ nonlinear LZ tunneling \cite{zhang2019} and deformation of
interference patterns \cite{li2018}, which will be left for furture
investigation. The St\"{u}ckelberg interferometer can still be implemented at
a small interaction energy per site ($<<$kHz), which can be tuned by means of
Feshbach resonance \cite{rb87}.

Besides the Stokes phase accumulated at the LZ transitions and the dynamical
phase accumulated during adiabatical evolution between the LZ transitions,
St\"{u}ckelberg interference also depends on a noncyclic geometric phase. This
noncyclic geometric phase is nonvanishing in special energy spectra
configuration such as those with Dirac cones \cite{lim2015}. The geometric
phase can also be made gauge-dependent, for example as proposed in a
periodically driven two-level system \cite{gasparinetti2011,kim2014}. A recent
experiment have also tested theory for the noncyclic geometric phase
\cite{zhouSA2020}. It would be interesting to consider geometric
St\"{u}ckelberg interferometer in future work and extend the discussion into
the non-Hermitian case \cite{wu2019}.

In summary, we have shown that SO-coupled cold atoms trapped in an optical
lattice can be used to implement St\"{u}ckelberg interference. It represents
an atom interferometry with synthetic gauge fields and provides new
opportunities to measure the synthetic gauge field. Time-dependent and
time-averaged spin probability is derived using Floquet-Bloch theory. Based on
that the interference patterns are computed in the parameter space directly
accessible in experiments and resonances are found. Finally we studied chiral
Bloch oscillation and found that the system can display a rich spin-mementum
locking via varying the detuning. The phenomena predicted in this work can be
readily observed in current available experiments on atomic flux lattices.

\begin{acknowledgments}
We thank Yongping Zhang for helpful discussions. This work is supported by
National Natural Science Foundation of China (Grants No. 12074120, No. 11374003, No.
11774093), the National Key Research and Development Program of China
(Grant No. 2016YFA0302001) and the Science and Technology Commission of Shanghai Municipality (Grant No. 20ZR1418500).
\end{acknowledgments}

\end{document}